\documentclass[journal]{IEEEtran}
\ifCLASSINFOpdf
\else
\fi
\pdfoutput=1
\usepackage[noend]{algpseudocode}
\usepackage{algorithmicx,algorithm}
\usepackage{cite}
\usepackage{multirow}
\usepackage{amsmath}  
\usepackage{float}
\usepackage{booktabs} 

\usepackage{graphicx}
\usepackage[colorlinks,linkcolor=black]{hyperref}
\usepackage{subfigure}
\usepackage{bm}
\usepackage{lineno}
\usepackage{url}
\usepackage{amsmath, amsfonts}
\usepackage{tikz}

\usepackage{dsfont}
\usepackage{verbatim}
\usetikzlibrary{positioning}
\usepackage{calc}

\begin{document}
\flushbottom
%
\title{TS-RNN: Text Steganalysis Based on Recurrent Neural Networks}
%
%
%

\author{Zhongliang~Yang,
        Ke~Wang,
        Jian~Li,
        Yongfeng~Huang,
        and~Yu-Jin~Zhang
\thanks{Z. Yang, K. Wang, J. Li, Y. Huang and Y. Zhang are with the Department of Electronic Engineering, 
Tsinghua University, Beijing, 100084, China. E-mail: yfhuang@tsinghua.edu.cn}}

\maketitle

\begin{abstract}

With the rapid development of natural language processing technologies, more and more text steganographic methods based on automatic text generation technology have appeared in recent years. These models use the powerful self-learning and feature extraction ability of the neural networks to learn the feature expression of massive normal texts. Then they can automatically generate dense steganographic texts which conform to such statistical distribution based on the learned statistical patterns. In this paper, we observe that the conditional probability distribution of each word in the automatically generated steganographic texts will be distorted after embedded with hidden information. We use Recurrent Neural Networks (RNNs) to extract these feature distribution differences and then classify those features into cover text and stego text categories. Experimental results show that the proposed model can achieve high detection accuracy. Besides, the proposed model can even make use of the subtle differences of the feature distribution of texts to estimate the amount of hidden information embedded in the generated steganographic text.

\end{abstract}

\begin{IEEEkeywords}
Text Steganography, Text Steganalysis, Recurrent Neural Network, Embedded Rate Estimation.
\end{IEEEkeywords}

%
\IEEEpeerreviewmaketitle

\section{Introduction}
%
%
%
%

As concluded by Claude E. Shannon \cite{shannon1949communication}, there are three main types of information security models in cyberspace: encryption system, privacy system, and concealment system. The concealment system is different from other two systems. It mainly protects information security by embedding the secret information into the common carrier, hidding the existence of confidential information to achieve the purpose of not being easily suspected and attacked \cite{Simmons1984The}. For information hidding, media with various forms, including image \cite{fridrich2009steganography,chen2019defining}, audio \cite{yang2017sudoku,tian2015optimal}, text \cite{xiang2014research,xiang2018linguistic,xiang2017novel}, can be adopted as carrier. As the most common and frequently used information interaction carrier in people's daily lives, using text as carrier to hide and transmit secret information has very important value and significance. Therefore, text steganography has attracted wide attention from researchers in recent years\cite{Luo2016Text,Luo2017Text,yang2018rits,yang2018rnn,yang2018markov}. 

In recent years, with the rapid development of deep learning in the field of Natural Language Processing, plenty of works related to high-quality readable text generation have appeared, including neural machine translation \cite{Bahdanau2014Neural}, dialogue systems \cite{Shang2015Neural}, and image caption \cite{yang2017image}. Based on these works, growing interests have been directed to text steganography by cover synthesis \cite{Luo2017Text,fang2017generating,yang2018rnn,yang2018rits}. These methods utilize the powerful feature extraction capabilities of neural networks, analyze the statistical feature distribution of a large number of training texts, and then reconstruct samples that conform to such statistical distribution. Furthermore, based on the generated texts format, there are two different types of methods: natural text steganography \cite{fang2017generating,yang2018rnn,yang2018rits} and special format text steganography \cite{Luo2017Text,Luo2016Text}. The difference is that in the process of generating steganographic texts, in addition to using the learned statistical language model, the special format text steganographic methods also combine some syntactic rules.

Generally, text steganalysis models mainly attempt to analyze and extract some text features first, and then they analyze the differences between these features before and after steganography to determine whether the text contains secret information. Traditional text steganalysis models mainly rely on some simple text features extracted artificially, such as word frequency \cite{Yang2010Linguistic}. However, the features they extracted and analyzed were very simple, which made them difficult to deal with the latest steganographic text generation methods based on neural networks. Recently, some researchers have tried to analyze the high-level semantic relationship between words in texts to determine whether the text contains hidden information\cite{yang2018ts,wen2019convolutional,8653856}.

In this paper, we first analysis that the conditional probability distribution of each word in the automatically generated steganographic texts will be distorted after embedded with hidden information. Then we propose a new text steganalysis mothod based on Recurrent Neural Networks (RNNs), which can extract the conditional probability features of each word in texts. Based on these features, we can then achieve satisfactory steganalysis performance and can even estimate the amount of hidden information contained in the text.

\section{TS-RNN Methodology}

For a text $X$ of length $n$, we can model it as a sequence signal with length $n$, that is, $X = \{x_1, x_2, ..., x_n\}$, where $x_i$ represents the $i$-th word. We hypothesize that once we embed hidden information in the text generation process, it is equivalent to superimposing noise on these conditional probability distributions, which will inevitably affect the probability distribution of the entire text, namely:

\begin{equation}
\begin{aligned}
p(\overline{X}) & = p(X) + p(\Delta)\\
& = p(\overline{x_1},\overline{x_2},\overline{x_3},...,\overline{x_n})\\
& = p(\overline{x_1})p(\overline{x_2}\mid \overline{x_1})...p(\overline{x_n}\mid \overline{x_1},\overline{x_2},...,\overline{x_{n-1}})\\
& = [p(x_1)+p(\delta_1)]\cdot[p(x_2\mid x_1)+p(\delta_2)]\\
& ...[p(x_n\mid x_1,x_2,...,x_{n-1})+p(\delta_n)].
\end{aligned}
\end{equation}

\noindent Here, $\overline{X}$ represents the generated steganographic text and $\overline{x_i}$ represents the $i$-th word in it. $\delta_i$ represents the disturbance caused by the embedded information on the conditional probability distribution of the $i$-th word. Therefore, our core idea is to realize the recognition of steganographic text by analyzing the difference in statistical distribution of texts caused by embedded hidden information.

The biggest characteristics of Recurrent Neural Network (RNN) is the feedback connections which makes it very suitable for modeling sequential signals. To avoid gradient vanish problem \cite{Hochreiter1998The}, we usually use Long Short-Term Memory (LSTM) \cite{Hochreiter1997Long} as the hidden units. An LSTM unit can be described using the following formulas:

\begin{equation}
 \left\{\begin{array}{l}
 I_t = \sigma(W_i \cdot [h_{t-1},x_t] + b_i), \\
 F_t = \sigma(W_f \cdot [h_{t-1},x_t] + b_f),\\
 C_t = F_t \cdot C_{t-1} + I_t \cdot \tanh(W_c \cdot [h_{t-1},x_t] + b_c),\\
 O_t = \sigma(W_o \cdot [h_{t-1},x_t] + b_o),\\
 h_t = O_t \cdot \tanh(C_t).\\
             \end{array}  
        \right.
\end{equation}

\noindent For simplicity, we denote the transfer function of LSTM units by $f_{LSTM}(\ast)$. 

For feature extraction, we first map each word to a dense semantic space with a dimension of $d$, that is $x_i \in \mathbb{R}^{d}$. Then, for each sentence $X$, we can illustrate it with a matrix $X \in \mathbb{R}^{L \times d}$, where the $i$-th row indicates the $i$-th word in sentence $X$ and $L$ is the length of it, that is

\begin{equation}
X = 
\left[                 
  \begin{array}{c}   
    x_{1}\\  
    x_{2}\\  
    \vdots\\
    x_{L}\\
  \end{array}
\right]
=
\left[                 
  \begin{array}{cccc}   
    a_{1,1} & a_{1,2} & \cdots\ & a_{1,d}\\  
    a_{2,1} & a_{2,2} & \cdots\ & a_{2,d}\\  
    \vdots & \vdots & \ddots & \vdots\\
    a_{L,1} & a_{L,2} & \cdots\ & a_{L,d}\\
  \end{array}
\right].
\end{equation}

In general, a recurrent neural network consists multiple network layers, each of which has multiple LSTM units. We use $n_j$ to indicate the LSTM units number of $j$-th hidden layer $U^j$, so the units of $j$-th layer can be represented as

$$U^j = \{u^j_{1},u^j_{2},\cdots,u^j_{n_j}\}.$$

The input of each unit $u^1_{i}$ at time step $t$ is the weighted sum of the elements in $x_t$, then the output value of $u^1_{i}$ at time step $t$ is

\begin{equation}
o^1_{i,t} = f_{LSTM}(u^1_{i,t}) = f_{LSTM}(\sum_{k=1}^dw^1_{i,k}\cdot a_{j,k} + b^1_{i,t}).
\end{equation}

\noindent Where $w^1_.$ and $b^1.$ are learned weights and biases, respectively.


Previous works have shown that within a certain range, the more layers of neural network in space, the stronger the ability to extract and express features \cite{Zeiler2013Visualizing}. So we stack the network with multiple layers of LSTM units and the transfer matrix between $l$-th layer and $(l+1)$-th layer can be represented as a matrix $W^{l} \in \mathbb{R}^{{n_l} \times {n_{l+1}}}$, that is 

\begin{equation}
u^l_{i,t} = O^{l-1}_t \cdot W^{l,m} = \sum_{k=1}^{n_{l-1}}w^l_{i,k}\cdot o^{l-1}_{k,t} + b^l_{i,t}.
\end{equation}

And the output of the $l$-th layer at time step $t$ is

\begin{equation}
\begin{aligned}
&o^l_{i,t} = f_{LSTM}(u^l_{i,t}) = f_{LSTM}(\sum_{k=1}^{n_{l+1}}w^{l}_{i,k}\cdot o^l_{k,t} + b^{l}_{i,t}).
\end{aligned}
\end{equation}

According to Equation $(2)$, the output of the $l$-th hidden layer at time step $t$ can be regarded as a summary of all previous $t$ words $\{x_1,x_2,...,x_{t-1}\}$. 

\begin{table*}[!tp]

\caption{Results of different steganalysis methods on special format text set.}
\label{tab:4}
\centering

  \begin{tabular}{c|c|c|ccc|ccc|ccc|ccc|ccc}
  \toprule[2pt]
  \multicolumn{2}{c|}{Format} &bpw &\multicolumn{3}{|c|}{\cite{meng2009linguistic}} &\multicolumn{3}{|c|}{\cite{samanta2016real}}  &\multicolumn{3}{|c|}{\cite{din2015performance}} &\multicolumn{3}{|c|}{TS-RNN} &\multicolumn{3}{|c}{TS-BiRNN}\\
  \hline
  \multicolumn{3}{c|}{Metric} &Acc &P &R &Acc &P &R &Acc &P &R &Acc &P &R &Acc &P &R\\
  \hline
  \multirow{10}{*}{FW} &\multirow{5}{*}{FL} &1 &0.521 &0.597 &0.518 &0.768 &0.775 &0.768 &0.724 &0.706 &0.766 &\textbf{0.874} &0.854 &\textbf{0.902} &0.873 &\textbf{0.877} &0.868\\
  &  &2 &0.600 &0.671 &0.588 &0.868 &0.877 &0.868 &0.843 &0.840 &0.848 &\textbf{0.968} &\textbf{0.974} &\textbf{0.962} &0.964 &0.966 &\textbf{0.962}\\
  &  &3 &0.681 &0.747 &0.659 &0.884 &0.896 &0.884 &0.895 &0.896 &0.894 &0.966 &0.964 &0.968 &\textbf{0.971} &\textbf{0.970} &\textbf{0.972}\\
  &  &4 &0.705 &0.769 &0.681 &0.872 &0.885 &0.861 &0.899 &0.892 &0.908 &0.989 &0.986 &0.992 &\textbf{0.993} &\textbf{0.992} &\textbf{0.994}\\
  &  &5 &0.798 &0.858 &0.765 &0.868 &0.893 &0.864 &0.933 &0.930 &0.936 &0.988 &\textbf{0.992} &0.984 &\textbf{0.990} &\textbf{0.992} &\textbf{0.988}\\
  \cline{2-18}
  &\multirow{5}{*}{EL} &1 &0.515 &0.598 &0.513 &0.775 &0.775 &\textbf{0.775} &0.750 &0.778 &0.700 &0.751 &0.752 &0.747 &\textbf{0.778} &\textbf{0.793} &0.752\\
  &  &2 &0.592 &0.671 &0.579 &0.917 &0.956 &0.897 &0.904 &0.902 &0.906 &0.980 &0.982 &0.978 &\textbf{0.986} &\textbf{0.986} &\textbf{0.986}\\
  &  &3 &0.675 &0.742 &0.654 &0.918 &0.924 &0.918 &0.944 &0.932 &0.958 &0.990 &\textbf{0.996} &0.984 &\textbf{0.992} &0.992 &\textbf{0.992}\\
  &  &4 &0.712 &0.772 &0.689 &0.915 &0.948 &0.915 &0.961 &0.964 &0.958 &0.997 &0.996 &0.998 &\textbf{1.000} &\textbf{1.000} &\textbf{1.000}\\
  &  &5 &0.819 &0.869 &0.789 &0.911 &0.923 &0.901 &0.970 &0.974 &0.966 &\textbf{0.998} &\textbf{0.998} &\textbf{0.998} &0.996 &0.996 &0.996\\
  \toprule[1.2pt]
  \multirow{10}{*}{SW} &\multirow{5}{*}{FL} &1 &0.539 &0.616 &0.533 &0.666 &0.667 &0.666 &0.710 &0.703 &0.726 &0.699 &0.726 &0.637 &\textbf{0.739} &\textbf{0.744} &\textbf{0.727}\\
  &  &2 &0.619 &0.691 &0.604 &0.898 &0.901 &0.898 &0.918 &0.927 &0.908 &0.970 &0.961 &0.980 &\textbf{0.973} &\textbf{0.963} &\textbf{0.984}\\
  &  &3 &0.691 &0.762 &0.667 &0.928 &0.931 &0.941 &0.954 &0.960 &0.948 &0.979 &0.974 &\textbf{0.984} &\textbf{0.981} &\textbf{0.984} &\textbf{0.978}\\
  &  &4 &0.730 &0.796 &0.703 &0.925 &0.930 &0.898 &0.973 &0.972 &0.974 &0.990 &0.994 &0.986 &\textbf{0.994} &\textbf{0.996} &\textbf{0.992}\\
  &  &5 &0.810 &0.865 &0.779 &0.916 &0.927 &0.913 &0.985 &0.986 &0.984 &0.992 &0.994 &0.990 &\textbf{0.997} &\textbf{0.996} &\textbf{0.998}\\
  \cline{2-18}
  &\multirow{5}{*}{EL} &1 &0.523 &0.624 &0.519 &0.656 &0.656 &0.656 &0.675 &0.659 &\textbf{0.724} &0.695 &0.709 &0.659 &\textbf{0.722} &\textbf{0.736} &0.691\\
  &  &2 &0.589 &0.682 &0.575 &0.926 &0.927 &0.926 &0.940 &0.956 &0.922 &0.985 &\textbf{0.980} &0.990 &\textbf{0.987} &\textbf{0.980} &\textbf{0.994}\\
  &  &3 &0.651 &0.721 &0.633 &0.942 &0.942 &0.942 &0.950 &0.957 &0.942 &0.991 &\textbf{0.990} &0.992 &\textbf{0.994} &\textbf{0.990} &\textbf{0.998}\\
  &  &4 &0.696 &0.754 &0.675 &0.958 &0.959 &0.958 &0.978 &0.982 &0.974 &0.999 &0.998 &\textbf{1.000} &\textbf{1.000} &\textbf{1.000} &\textbf{1.000}\\
  &  &5 &0.793 &0.829 &0.774 &0.954 &0.957 &0.952 &0.984 &0.984 &0.984 &\textbf{0.998} &\textbf{0.998} &\textbf{0.998} &0.994 &0.996 &0.992\\
  \bottomrule[2pt] 
  \end{tabular}
\end{table*}

\begin{figure}[!tp]
\centering
\includegraphics[width=\linewidth]{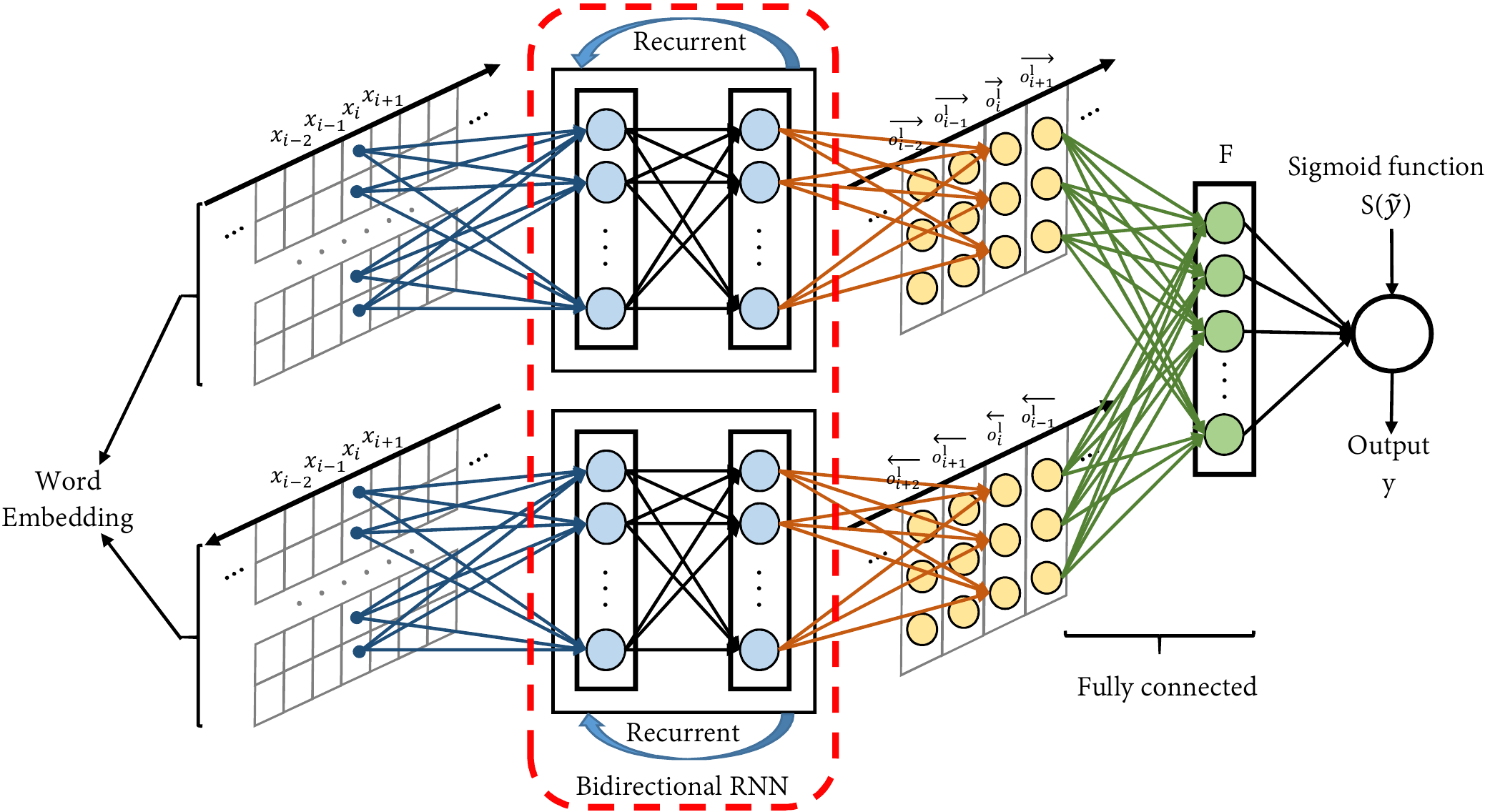}
\caption{We use bidirectional recurrent neural network (BiRNN) to extract the potential correlation of each word to all surrounding words, and then we use these extracted correlation features to classify whether the original text has hidden data.}
\label{fig:6}
\end{figure}

In order to further extract the potential correlations of each word to all surrounding words (including the previous words and the following words), we further added a reverse RNN, which has been shown in Figure \ref{fig:6}. The forward RNN is more inclined to extract the correlation between each word and the previous words, while the reverse RNN is mainly focused on extracting the correlation between each word and the following words.

We can use $\overrightarrow{O^l}$ and $\overleftarrow{O^l}$ to represent the correlation features of the words extracted by the forward RNN and the backward RNN, respectively. In order to fuse these two features, we first take out the feature vector of their last moment and splice them together. The spliced vector is named as $Z$:

\begin{equation}
Z = [z_1,z_2,...,z_m] = [\overrightarrow{O^l};\overleftarrow{O^l}].
\end{equation}

Then we define a Feature Fusion Matrix $W^{F} \in \mathbb{R}^{{m} \times {h}}$ to fuse the features, where $h$ is the dimension of the fused feature vector $F = [f_1,f_2,...,f_h]$, where:

\begin{equation}
f_{j} = \sum^m_{i=1}w^F_{i,j}\cdot z_i + b^F_{i}.
\end{equation}

Finally, following previous works \cite{yang2019real,yang2018clinical}, we pass the feature vector $F$ through the softmax classifier $S$ and the output is:

\begin{equation}
y = S(\widetilde{y}) = S(\sum^{h}_{i=1}f_{i}\cdot c_{i}+b^o_i).
\end{equation}

The output value reflects the probability that our model believes that the text contains covert information. We can then set a detection threshold and then the final detection result can be expressed as

\begin{equation}
X \in 
\left\{
\begin{aligned}
& Stegotext & (y \geq threshold) \\
& Covertext & (y < threshold)
\end{aligned}
\right.
\end{equation}

\noindent In other words, the model tries to predict the label (0 for normal, 1 for stego) for a given text.

In the process of training, we update network parameters by applying backpropagation algorithm, and the loss function of the whole network consists of two parts, one is the error term and the other is the regularization term, which can be described as:

\begin{equation}
LOSS = - \frac{1}{N}\sum_{N}T_i \cdot log(Y_i) + \|C\|_2,
\end{equation}

\noindent where $N$ is the batch size of texts. $Y_i$ represents the probability that the $i$-th sample is judged to contain covert information, $T_i$ is the actual label of the $i$-th sample. In order to strengthen the regularization and prevent overfitting, we adopt the dropout mechanism \cite{krizhevsky2012imagenet}.

\begin{table*}[!tp]
\caption{Results of different steganalysis methods on natural text set.}
\label{tab:5}
\centering
  \begin{tabular}{c|c|ccc|ccc|ccc|ccc|ccc}
  \toprule[2pt]
  \multicolumn{2}{c|}{Method} &\multicolumn{3}{|c|}{\cite{meng2009linguistic}} &\multicolumn{3}{|c|}{\cite{samanta2016real}}  &\multicolumn{3}{|c|}{\cite{din2015performance}} &\multicolumn{3}{|c}{TS-RNN} &\multicolumn{3}{|c}{TS-BiRNN}\\
  \hline
  Format &bpw &Acc &P &R &Acc &P &R &Acc &P &R &Acc &P &R &Acc &P &R\\
  \hline
  \multirow{5}{*}{News} &1 &0.532 &0.517 &0.382 &0.763 &0.739 &0.812 &0.840 &0.869 &0.801 &0.911 &0.914 &0.909 &\textbf{0.915} &\textbf{0.915} &\textbf{0.914}\\
  &2 &0.513 &0.535 &0.204 &0.786 &0.762 &0.832 &0.835 &0.867 &0.791 &0.917 &\textbf{0.937} &0.895 &\textbf{0.924} &0.919 &\textbf{0.931}\\
  &3 &0.597 &0.679 &0.367 &0.824 &0.767 &0.931 &0.897 &0.909 &0.882 &0.965 &0.955 &0.975 &\textbf{0.968} &\textbf{0.958} &\textbf{0.979}\\
  &4 &0.755 &0.831 &0.640 &0.859 &0.797 &0.962 &0.938 &0.962 &0.911 &0.972 &\textbf{0.972} &\textbf{0.972} &\textbf{0.974} &0.967 &\textbf{0.982}\\
  &5 &0.847 &0.918 &0.761 &0.881 &0.829 &0.959 &0.961 &0.976 &0.945 &\textbf{0.991} &\textbf{0.988} &\textbf{0.994} &0.990 &0.987 &\textbf{0.994}\\
  \hline
  \multirow{5}{*}{IMDB} &1 &0.577 &0.642 &0.345 &0.767 &0.779 &0.744 &0.787 &0.829 &0.722 &0.906 &0.951 &\textbf{0.856} &\textbf{0.910} &\textbf{0.960} &0.855\\
  &2 &0.713 &0.807 &0.560 &0.849 &0.934 &0.871 &0.869 &0.911 &0.818 &\textbf{0.964} &\textbf{0.982} &\textbf{0.946} &0.963 &0.980 &\textbf{0.946}\\
  &3 &0.840 &0.925 &0.741 &0.900 &0.877 &0.931 &0.916 &0.944 &0.885 &0.970 &0.981 &0.959 &\textbf{0.973} &\textbf{0.984} &\textbf{0.962}\\
  &4 &0.909 &0.969 &0.845 &0.937 &0.905 &0.975 &0.962 &0.975 &0.947 &0.989 &0.992 &0.987 &\textbf{0.992} &\textbf{0.996} &\textbf{0.989}\\
  &5 &0.909 &0.989 &0.828 &0.929 &0.921 &0.940 &0.977 &0.987 &0.966 &\textbf{0.996} &\textbf{0.998} &\textbf{0.994} &\textbf{0.996} &\textbf{0.998} &\textbf{0.994}\\
  \hline
  \multirow{5}{*}{Twitter} &1 &0.538 &0.520 &0.387 &0.654 &0.652 &0.658 &0.665 &0.664 &0.670 &\textbf{0.801} &\textbf{0.836} &0.749 &0.791 &0.806 &\textbf{0.767}\\
  &2 &0.544 &0.523 &0.399 &0.745 &0.762 &0.712 &0.750 &0.827 &0.631 &0.849 &0.887 &\textbf{0.800} &\textbf{0.850} &\textbf{0.895} &0.794\\
  &3 &0.577 &0.669 &0.303 &0.809 &0.798 &0.826 &0.834 &0.889 &0.764 &0.916 &0.936 &\textbf{0.894} &\textbf{0.924} &\textbf{0.951} &0.893\\
  &4 &0.729 &0.836 &0.570 &0.842 &0.824 &0.871 &0.885 &0.950 &0.813 &\textbf{0.945} &\textbf{0.966} &0.921 &0.939 &0.923 &\textbf{0.958}\\
  &5 &0.850 &0.916 &0.770 &0.851 &0.839 &0.870 &0.899 &\textbf{0.961} &0.832 &0.940 &0.939 &\textbf{0.942} &\textbf{0.943} &0.946 &0.939\\
  \bottomrule[2pt] 
  \end{tabular}
\end{table*}

\begin{figure*}[!tp]
  \centering
  \includegraphics[width=0.32\linewidth]{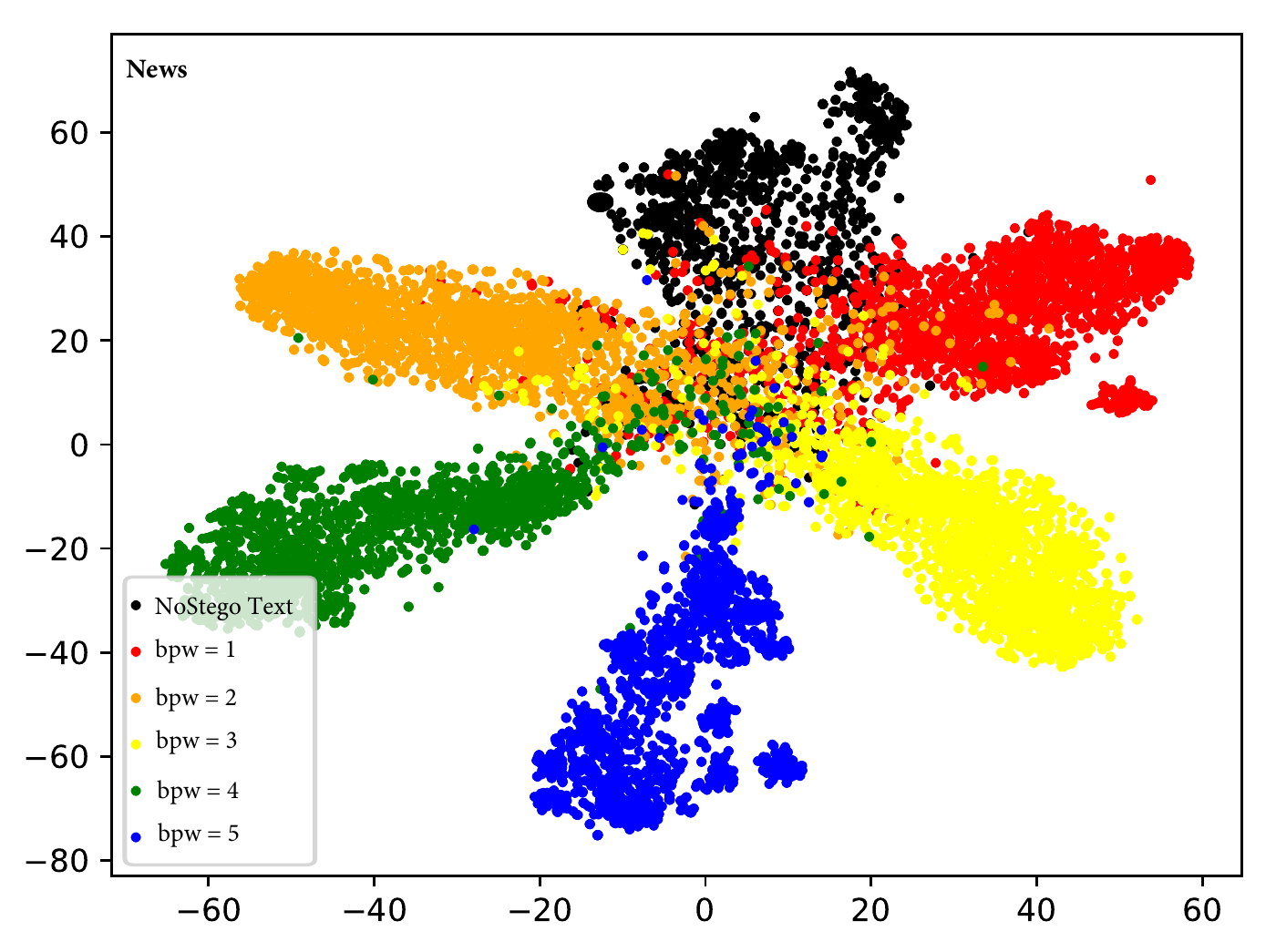}
  \includegraphics[width=0.32\linewidth]{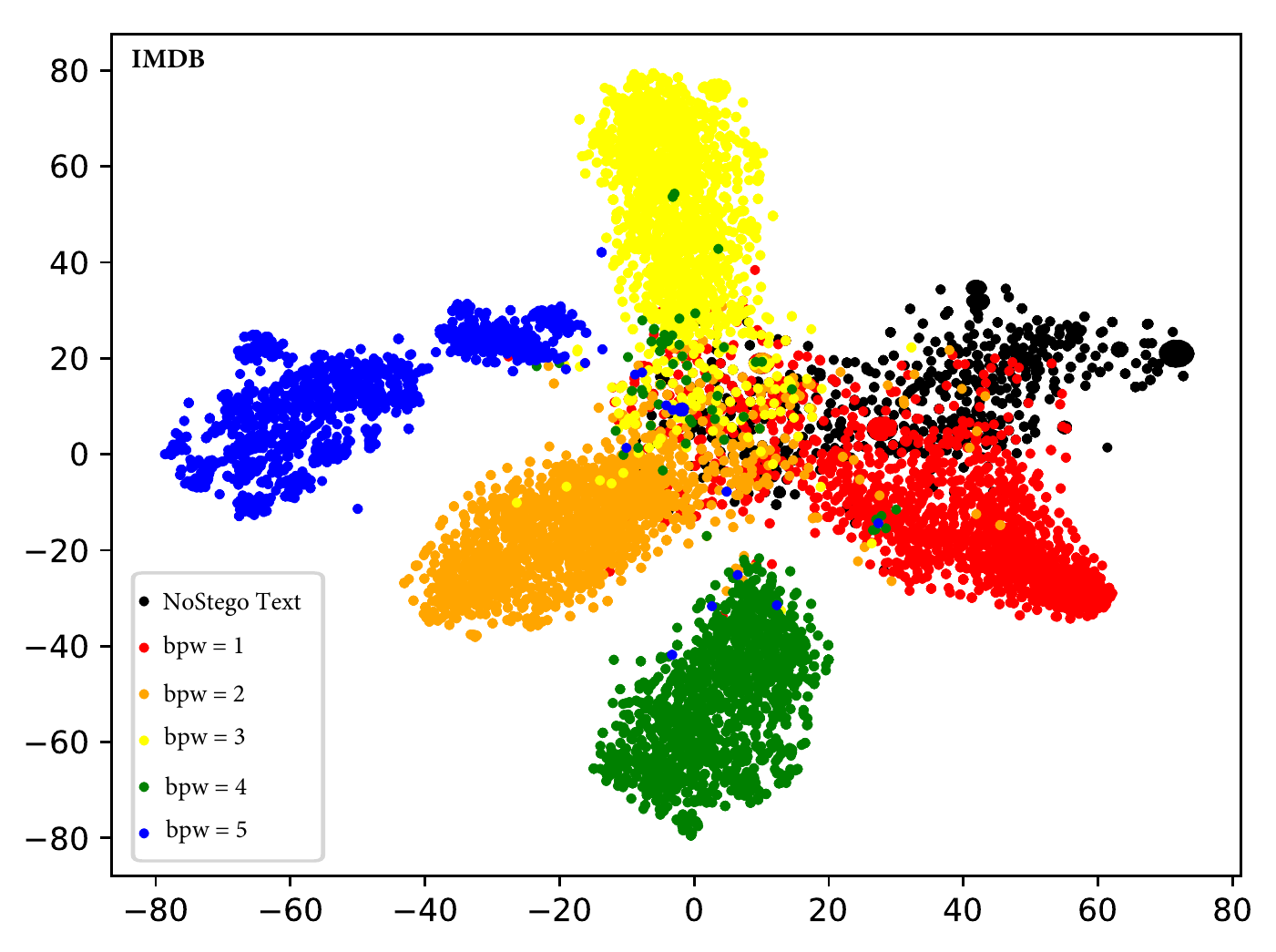}
  \includegraphics[width=0.32\linewidth]{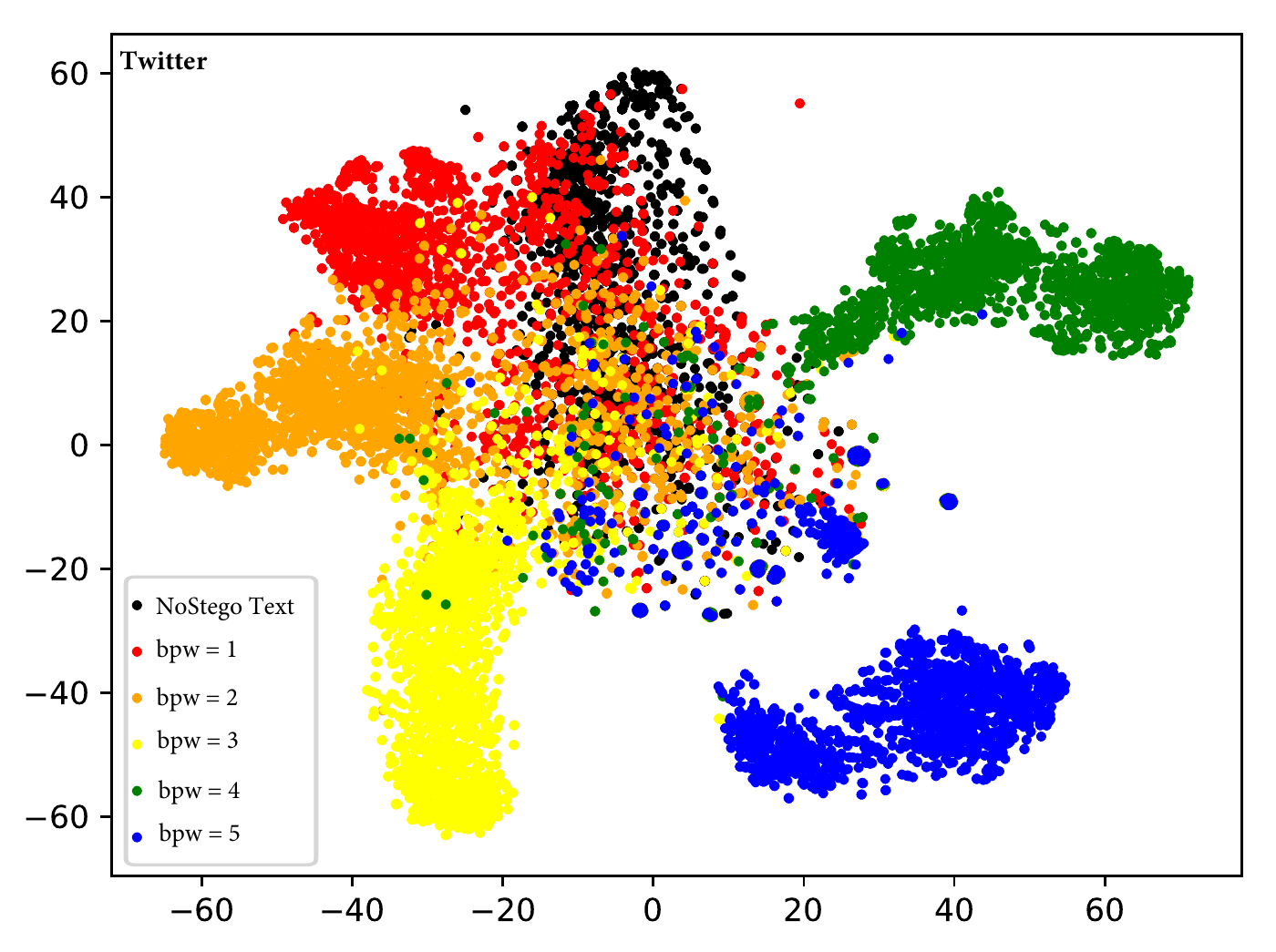}
  \caption{The distribution of correlations features under different information embedding rates in the feature space.}
\label{fig:10}
\end{figure*}

\begin{table*}[ht]

\caption{The results of the proposed models' estimate of the capacity of the covert information in text.}
\label{tab:6}
\centering
  \begin{tabular}{c|c|ccc|ccc|ccc|ccc|ccc}
  \toprule[1.5pt]
  \multicolumn{2}{c|}{Format} &\multicolumn{3}{|c|}{\cite{meng2009linguistic}} &\multicolumn{3}{|c|}{\cite{samanta2016real}}  &\multicolumn{3}{|c|}{\cite{din2015performance}} &\multicolumn{3}{|c|}{TS-RNN} &\multicolumn{3}{|c}{TS-BiRNN}\\
  \hline
  \multicolumn{2}{c|}{Metric} &P &R &$F_1$ &P &R &$F_1$ &P &R &$F_1$ &P &R &$F_1$ &P &R &$F_1$\\
  \hline
  \multirow{2}{*}{FW} &\multirow{1}{*}{FL} &0.258 &0.297 &0.245 &0.465 &0.473 &0.453 &0.465 &0.472 &0.476 &0.699 &0.698 &0.698 &\textbf{0.724} &\textbf{0.722} &\textbf{0.722}\\
  \cline{2-17}
  &\multirow{1}{*}{EL} &0.266 &0.311 &0.258 &0.510 &0.511 &0.503 &0.513 &0.515 &0.514 &0.692 &0.688 &0.685 &\textbf{0.738} &\textbf{0.734} &\textbf{0.732}\\
  \hline
  \multirow{2}{*}{SW} &\multirow{1}{*}{FL} &0.261 &0.292 &0.253 &0.493 &0.492 &0.485 &0.496 &0.501 &0.498 &0.679 &0.674 &0.674 &\textbf{0.700} &\textbf{0.699} &\textbf{0.697}\\
  \cline{2-17}
  &\multirow{1}{*}{EL} &0.246 &0.286 &0.239 &0.540 &0.521 &0.533 &0.568 &0.568 &0.566 &0.694 &0.688 &0.685 &\textbf{0.726} &\textbf{0.725} &\textbf{0.724}\\
  \hline
  \hline
  \multicolumn{2}{c|}{News} &0.445 &0.396 &0.420 &0.701 &0.706 &0.703 &0.745 &0.741 &0.743 &0.905 &0.904 &0.904 &\textbf{0.908} &\textbf{0.908} &\textbf{0.908}\\
  \hline
  \multicolumn{2}{c|}{IMDB} &0.490 &0.512 &0.501 &0.742 &0.745 &0.743 &0.767 &0.760 &0.763 &0.917 &0.914 &0.914 &\textbf{0.929} &\textbf{0.927} &\textbf{0.927}\\
  \hline
  \multicolumn{2}{c|}{Twitter} &0.417 &0.363 &0.303 &0.620 &0.620 &0.620 &0.638 &0.615 &0.626 &0.800 &0.797 &0.798 &\textbf{0.806} &\textbf{0.801} &\textbf{0.803}\\
  \bottomrule[1.5pt] 
  \end{tabular}
\end{table*}

\section{Experiments and Analysis}

We trained our model by using the T-Steg dataset \footnote{\url{https://github.com/YangzlTHU/TS-CNN}} which was released by Z. Yang \emph{et al.} \cite{yang2018ts}. The T-Steg dataset contains two categories of steganographic texts: special format steganographic texts (Chinese) generated by the model proposed by Luo \emph{et al.} \cite{Luo2017Text}, and natural steganographic texts (English) generated by the model proposed by Fang \emph{et al.} \cite{fang2017generating}. In the T-Steg dataset, two different forms of steganographic Chinese poetries are provided: poetries with five words (FW) per line and poetries with seven words (SW) per line. Each format can be further divided into two categories, that is, each poem contains four lines (FL) or eight lines (EL). For natural texts, T-Steg contains the most common text media on the Internet, including Twitter, movie reviews and News. Both steganographic methods can generate steganographic texts with different embedding rate by altering the number of bits hidden in per word (bpw). In T-Steg dataset, it contains 10,000 steganographic sentences for different types of text with different embedding rates.

The hyper-parameters in the proposed model were finally determined based on the comparison experiment: we mapped each word to a 256-dimensions vector. The number of hidden layers was $3$ in TS-RNN and $2$ in TS-BiRNN. The number of LSTM units per layer was setted to be $200$ for TS-RNN and $100$ for TS-BiRNN. We setted the detection threshold to be $0.5$. We used $tanh$ as nonlinear activation function $\sigma$ in Equation $(6)$. We chose Adam \cite{Kingma2014Adam} as the optimization method. The learning rates were initially setted as 0.001 and batch size was setted as 128, dropout rate was 0.5. 

\subsection{Evaluation Results and Discussion}

\subsubsection{Steganalysis Accuracy}

In this section, we compaired with three representative text steganalysis algorithms, which are proposed in \cite{meng2009linguistic}, \cite{samanta2016real}, \cite{din2015performance}, respectively. We use several evaluation indicators commonly used in classification tasks to evaluate the performance of our model, which are precision, recall, F1-score and accuracy. Experiment results have been shown in Table \ref{tab:4} and Table \ref{tab:5}.

According to the results, we can draw the following conclusions. Firstly, compared to other text steganalysis methods, the proposed models has achieved the best detection results on various metrics, including different text format and different embedding rates. Secondly, in Table \ref{tab:4} and Table \ref{tab:5}, we notice that in most cases, the detection performance of each model has improved with the increasing of the embedding rate. These results do meet our previous conjecture that with the increasing of embedding rate, it will damage the coherence of text semantics.

\subsubsection{Embedded Rate Estimation}

We can use t-Distributed Stochastic Neighbor Embedding (t-SNE) \cite{Maaten2014Accelerating} technique for the dimensionality reduction and visualization of this feature space, which can be found in Figure \ref{fig:10}. In this feature space, each point represents a sentence and different colors indicate different embedding rates. From Figure \ref{fig:10}, we can clearly see that as the embedded rate of hidden information increases in the generated texts, their distribution in the semantic space will gradually change. Further, based on these features, we find our model can estimate the capacity of hidden information inside. 

We mixed the texts at various embedding rates, i.e. $bpw=\{0,1,2,3,4,5\}$, and then we used different models to conduct multi-classification experiments. The experimental results are shown in Table \ref{tab:6}. From Table \ref{tab:6}, we can see that our model can achieve an estimated accuracy higher than 70\% and 90\% for the hidden information in the special format texts and natural texts, which outperforms all the other models. 

\section{Conclusion}

In this paper, we use bidirectional recurrent neural network (BiRNN) to extract the conditional distrubution features of each word in texts. Based on the distribution of these features, the proposed model achieves nearly 100\% precision and recall. Besides, the proposed model can even make use of the subtle distribution difference of the features to estimate the capacity of the hidden information inside, which shows state-of-the-art performance.



%

\section*{Acknowledgment}

This work was supported in part by the National Key Research and Development Program of China under Grant SQ2018YGX210002 and the National Natural Science Foundation of China (No.U1536207, No.U1705261 and No.U1636113).

\ifCLASSOPTIONcaptionsoff
  \newpage
\fi



\bibliographystyle{IEEEtran}
%
\bibliography{sample}

\end{document}